\documentclass[12pt, a4paper]{article}

\usepackage{graphicx}
\usepackage{bmpsize}
\usepackage{amsmath}
\usepackage{amsfonts}
\usepackage{amssymb}
\usepackage{latexsym}
\usepackage{graphicx}
\usepackage{color}
\usepackage{mathrsfs}
\usepackage{slashed}
\usepackage{soul}
\usepackage{array}
\usepackage{cite}
\usepackage{placeins}
\usepackage{hyperref}

\usepackage{enumerate} 
\usepackage{multirow}
\usepackage[table]{xcolor}
\usepackage{colortbl}
\usepackage{tikz-feynman}
\tikzfeynmanset{compat=1.0.0}

\usepackage{filecontents,catchfile}
\usepackage{filecontents,catchfile}

\usepackage{subcaption}

\usepackage{float}
\usepackage{braket}
\usepackage{xfrac}
\usepackage
[
version=4,
math-greek=default,
text-greek=default,
]{mhchem
}
\usepackage{booktabs}  

\usepackage[text={7in,10in}]{geometry}

\usepackage{comment}
\includecomment{toexclude} 
\excludecomment{addsections} 

\usepackage{pdflscape}

\usepackage[
  output-decimal-marker=.,
  separate-uncertainty=true,
]{siunitx}
\usepackage{color}

\usepackage{slashed}
\usepackage{xfrac}

\numberwithin{equation}{section}

\newcommand{\lag}{\mathcal L}
\newcommand{\sqt}{\frac{1}{\sqrt 2}}
\newcommand{\hlf}{\frac{1}{2}}
\newcommand{\hc}{\mathrm{h.c.}}

\begin{document}


\begin{center}
{\Huge \bf
Muon $g-2$ Anomaly from Vector-like Leptons in a 2-Higgs-doublet + Scalar Singlet Model
}
\\ [2.5cm]
{\large{\textsc{ 
Tim Brune\texorpdfstring{\textsuperscript{$\dagger$}}{\textsubscript{$\dagger$}}\footnote{\textsl{tim.brune@tu-dortmund.de}}
}}}
{\large{\textsc{ 
Thomas W. Kephart\texorpdfstring{\textsuperscript{*}}{\textsubscript{*}}\footnote{\textsl{thomas.w.kephart@vanderbilt.edu}}
}}}
{\large{\textsc{ 
Heinrich P\"as\texorpdfstring{\textsuperscript{$\dagger$}}{\textsubscript{$\dagger$}}\footnote{\textsl{heinrich.paes@tu-dortmund.de}}
}}}
\\[1cm]

\large{\textit{
\texorpdfstring{\textsuperscript{$\dagger$}}{\textsubscript{$\dagger$}}Fakult\"at f\"ur Physik, Technische Universit\"at Dortmund,\\
44221 Dortmund, Germany
\\
\texorpdfstring{\textsuperscript{*}}{\textsubscript{*}}Department of Physics and Astronomy, Vanderbilt University, Nashville, TN 37235
}}
\\ [2 cm]

{ \large{\textrm{
{\bf Abstract} 
}}}
\end{center}
\normalsize
We revisit an economical model for the $g-2$ anomaly featuring a vector-like charged fermion, a scalar doublet and a scalar singlet in the light of new results from Fermilab. 
The phenomenological implications for lepton-flavor universality, Higgs decays, and 
charged lepton-flavor violating decays are discussed in detail.

\def\thefootnote{\arabic{footnote}}
\setcounter{footnote}{0}
\pagestyle{empty}

\newpage
\pagestyle{plain}
\setcounter{page}{1}
\section{Introduction}
The 2021 
confirmation of the 20 years old muon g-2 anomaly by Fermilab's Muon g-2 Collaboration \cite{PhysRevLett.126.141801} 
and the subsequent runs 2 and 3 have inspired a new wave of theoretical work discussing physics beyond the Standard Model scenarios as an explanation (for an overview see for example \cite{Athron:2021iuf} and references therein). The most recent results \cite{Aguillard} indicate a $5.0 \sigma$ deviation from the current theoretical prediction \cite{Aoyama:2020ynm,aoyama:2012wk,Aoyama:2019ryr,czarnecki:2002nt,gnendiger:2013pva,davier:2017zfy,keshavarzi:2018mgv,colangelo:2018mtw,hoferichter:2019mqg,davier:2019can,keshavarzi:2019abf,kurz:2014wya,melnikov:2003xd,masjuan:2017tvw,Colangelo:2017fiz,hoferichter:2018kwz,gerardin:2019vio,bijnens:2019ghy,colangelo:2019uex,Blum:2019ugy,colangelo:2014qya},
\begin{align}
\Delta a_\mu = a_\mu^{exp}-a_\mu^{SM} = 2.49 \pm 0.48 \times 10^{-9}  \label{eq:amuexp}
\end{align}
with a precision of 0.20 ppm.
In this paper we revisit a model that we consider to be an economical and natural explanation of the anomaly \cite{Kephart_2002}. 
The scenario extends the Standard Model with a vector-like charged fermion, a scalar singlet and an additional Higgs doublet. While the vector-like charged fermion and the scalar singlet alone already induce a contribution to the muon anomalous magnetic moment, it is excluded as a solution for the anomaly due to the required large mixing angles between the SM Higgs doublet and the scalar singlet. These constraints can be evaded by invoking a second Higgs doublet.

All of the above mentioned BSM particles can arise naturally in 
$E_6$ unification \cite{Gursey:1975ki, Shafi:1978gg, Achiman:1978vg}, Trinification 
\cite{Rujula:1984,Babu:1985gi,Dvali:1994vj,Dvali:1994wj,Kephart:2001ix,Willenbrock:2003ca,Kim:2004pe,Sayre:2006ma,Kephart:2006zd,Cauet:2010ng,Hetzel:2015bla,Hetzel:2015cca,Pelaggi:2015kna,Babu:2021hef,Babu:2017xlu,Wang:2018yer,Raut:2022ryj}
and other 
product group models inspired for example by type IIB string theory compactified on $AdS_5/Z_n$~\cite{Kachru:1998ys,Frampton:1999wz,Frampton:1999zy,Frampton:2000mq}.  

In $E_6$ unification, the unified gauge group breaks down to the Standard Model via the
symmetry
breaking chain
$E_{6}$ $\rightarrow SO(10)\rightarrow SU(5)\rightarrow SU(3)\times
SU(2)%
\times U(1)$,
and an  $E_{6}$ family of fermions decomposes as:

\begin{eqnarray}{}
27\rightarrow 16+10+1\rightarrow
(\bar{5}+10+1)+(5+\bar{5})+1 \nonumber \\
\rightarrow {\rm(SM~family)} 
+(3,1)+(\bar{3},1)+(1,2)+(1,\bar{2})+2(1,1).
\end{eqnarray}

Likewise, in Trinification models, the Standard Model is embedded in the unifying $SU(3)_c \times SU(3)_L \times SU(3)_R$ gauge group
that is the maximal subgroup of $E_6$.
After spontaneous symmetry breaking, a  Trinification family $[(1,3,\bar{3})+(3,\bar{3},1)+(\bar{3},1,3)]$ then decomposes into 
\begin{equation}{}
[{\rm (SM~family)}+(3,1)+(\bar{3},1)+(1,2)+(1,\bar{2})+2(1,1)]\,,
\end{equation}
just like in $E_6$.
Here the conjugate pair of doublets $(1,2)+(1,\bar{2})$ describes a vector-like lepton. 
Combined with a scalar singlet and a second Higgs doublet that can easily arise in both schemes, they induce a new contribution to the
muon anomalous magnetic moment~\cite{Kephart_2002} that provides a natural and elegant
explanation for the g-2 anomaly (that is different from the one proposed in \cite{Raut:2022ryj} based on the gauge boson contribution).   
In the following discussion, we will not specify how the Lagrangian introduced in the next section can arise in such models and focus on the phenomenological implications of a given set of interactions.

This paper is organized as follows: In the next section, we discuss the field content of the model and construct it's 
scalar potential. We also study the vacuum structure and discuss the issue of vacuum stability. In section \ref{sec:muon}, we continue with the 
discussion of fermion mixing and the
calculation
of the muon anomalous magnetic moment and study its dependence on the free parameters of the model. In section \ref{sec:constraints}, we discuss the model's 
phenomenological implications for Higgs decays and lepton-flavor universality. In section \ref{sec:results}, we present data sets that are compatible with the experimental constraints and that allow us to explain the anomalous magnetic moment. A summary and conclusions are given in section \ref{sec:conclusion}.

\section{Model Overview}
\label{sec:model}
The model we consider consists of two Higgs doublets $H_{1,2}$ and a scalar singlet $S$ with expansions around the respective vacuum expectation values (vevs) $v_{1,2,S}$ given by
\begin{align}
    H_{1,2} = \begin{pmatrix} \phi_{1,2}^+ \\ \sqt(v_{1,2} + h_{1,2} + i \rho_{1,2}) \end{pmatrix}\,,\quad S = \sqt \left( v_S + S^\prime\right)\,,
\end{align}
while the relevant lepton sector consists of three $SU(2)_L$ doublets, 
\begin{align}
  L_\mu = \begin{pmatrix} \nu_\mu \\\mu\end{pmatrix}_{\!\!L}\,, \qquad L_M = \begin{pmatrix} \nu_M \\M\end{pmatrix}_{\!\!L}\,, \qquad L_M^\prime = \begin{pmatrix} \nu_M \\M\end{pmatrix}_{\!\!R}\,,
\end{align}
and a $SU(2)_L$ singlet $\mu_R$. \\
In order to reduce the complexity of the model, we impose a $Z_4$ symmetry with charge assignments given in Tab. \ref{tab:fields}. In particular, the $Z_4$ symmetry ensures the absence of flavour changing neutral currents and prevents the existence of terms with an odd number of scalar singlets, resulting in a scalar potential given by \footnote{See \cite{Branco:2011iw} for a detailed review on two Higgs doublet models. A recent extended discussion of the vacuum structure of 2HDMs, including numerous up to date 2HDM references, can be found in \cite{Ramsey-Musolf:2024zex}.} 
\begin{align}
\begin{split}
  V = &-\mu_1^2|H_1|^2-\mu_2^2|H_2|^2 - \mu_S^2 S^2 + \lambda_1|H_1|^4+ \lambda_2|H_2|^4 + \lambda_S S^4 \\
  &+2 \lambda_3 |H_1|^2|H_2|^2 +2 \lambda_4 (H_1^\dagger H_2)(H_2^\dagger H_1)   + \lambda_5\left[(H_1^\dagger H_2)^2 + (H_2^\dagger H_1)^2 \right] \\
  &+ 2\eta_1 |H_1|^2S^2+ 2\eta_2 |H_2|^2S^2\,. \label{eq:potential}
  \end{split}
\end{align}
In order to obtain a contribution to the muon anomalous magnetic moment, the $Z_4$ charges of the lepton sector are chosen such that $L_\mu$ mixes with the muon-type vectorlike fermions $L_M, L_M^\prime$ as 
\begin{align}
  \lag = -g \overline{L_\mu} H_1 \mu_R - y \overline{L_M} H_2 \mu_R - y^\prime \overline{L_\mu}L_M^\prime S - m_M \overline{L_M}L_M^\prime\,. \label{lag}
\end{align}
\begin{table}
  \centering
  \begin{tabular}{c c c c c c c c}
  \toprule
  {} & $H_1$& $H_2$ & $L_{\mu}$ & $\mu_R$ & $ L_M $& $ L_M^\prime $& $S$ \\
  \midrule 
  $\left( SU(3)_C,SU(2)_L\right)_{U(1)_Y}$ & $(1,2)_{\sfrac{1}{2}}$ & $(1,2)_{\sfrac{1}{2}}$ & $(1,2)_{-\sfrac{1}{2}}$ & $(1,1)_{-1}$ & $(1,2)_{-\sfrac{1}{2}}$ & $(1,2)_{-\sfrac{1}{2}}$ & $(1,1)_0$ \\
  $Z_4$ & $1$ & $-1$ & $-i$ & $-i$ & $+i$ & $+i$ & $-1$ \\
  \bottomrule
\end{tabular}
\caption{Field content and charge assignments relevant for the discussed model. }
\label{tab:fields}
\end{table}
A key ingredient to a sizable contribution to the muon anomalous magnetic moment is the mixing of the Higgs doublets $H_{1,2}$ with the scalar $S$ by means of the couplings $\eta_{1,2}$ in eq. \ref{eq:potential}. In order to simplify the model further, we will restrict the following analysis to the case $\eta_1 = 0, \eta_2 \neq 0$ which will prove to be sufficient to explain the muon data. 
\subsection{Scalar Sector}
\label{sec:scalar}
\subsubsection{Mass Spectrum and Alignment Limit}
The mass matrix for the CP-even scalars in the $(h_1,h_2,S)$ basis is given by 
\begin{align}
  M^2 = \begin{pmatrix} M^2_{h_1h_1} & M^2_{h_1h_2} & M^2_{h_1S} \\ M^2_{h_1h_2} & M^2_{h_2h_2} & M^2_{h_2S} \\ M^2_{h_1S} & M^2_{h_2S} & M^2_{SS}, \end{pmatrix}
\end{align}
with the matrix elements given as
\begin{align}
 M^2_{h_1h_1} &=   \frac{2 \lambda _1 \left(v^2 \tilde{\eta }_2+\kappa \right)}{\sigma }\,, \\
 M^2_{h_1h_2} &= 2 \lambda _{345} \sqrt{\frac{v^2 \tilde{\eta }_2+\kappa }{\sigma }} \sqrt{\frac{\lambda _{345} v^2 \lambda _S-\kappa }{\sigma }} \,,\\
 M^2_{h_1S} &= 0\,, \\
 M^2_{h_2h_2} &=  -\frac{2 \lambda _2 \left(\kappa -\lambda _{345} v^2 \lambda _S\right)}{\sigma }\,,\\
  M^2_{h_2S} &= 2 \eta _2 \sqrt{\frac{\lambda _{345} v^2 \lambda _S-\kappa }{\sigma }} \sqrt{\frac{\delta _{345} \eta _2-\Lambda _2 \mu _S^2}{\sigma }}\,, \\
  M^2_{SS} &=-\frac{2 \lambda _S \left(\Lambda _2 \mu _S^2+\eta _2 \left(\lambda _{345} v^2-\mu _2^2\right)\right)}{\sigma }. 
\end{align}
Here we introduced the shorthand notations
\begin{align}
  \lambda_{345} &= \lambda_{3}+\lambda_{4}+\lambda_{5}\,,&\quad
  \Lambda_i &= \lambda_i-\lambda_{345}\,,&\quad
  \sigma&=\eta_2^2-\Lambda_2\lambda_S\,,\nonumber\\
  \kappa&= \lambda_S\mu_2^2 - \eta_2\mu_S^2\,,&\quad
  \delta_i&=\mu_2^2-\lambda_i v^2\,,&\quad
  \tilde\eta_2&= \eta_2^2-\lambda_2\lambda_S\, \label{eq:abbreviations}
\end{align}
and $v= \sqrt{v_1^2 + v_2^2} =\SI{246}{\giga\electronvolt}$ is the Standard Model vev.
We obtain the mass eigenstates and eigenvalues via
\begin{align}
  \begin{pmatrix} h^0 \\ H^0 \\ S^0 \end{pmatrix} = U \begin{pmatrix} h_1 \\ h_2 \\ S^\prime \end{pmatrix}\,,  \qquad M^2_{diag}:=\mathrm{diag}(m^2_{h^0}, m^2_{H^0}, m^2_{S^0}) = U^T M^2 U \,,\label{eq:massbasis} 
\end{align}
where
\begin{align}
  U = \begin{pmatrix}  c_{12}c_{13} & s_{12}c_{13} & s_{13} \\ -s_{12}c_{23}-c_{12}s_{23}s_{13} & c_{12}c_{23}-s_{12}s_{23}s_{13} & s_{23}c_{13} \\ s_{12}s_{23}-c_{12}c_{23}s_{13} & - c_{12}s_{23} - s_{12}c_{23}s_{23} & c_{23}c_{12} \end{pmatrix}\,
\end{align}
with $c_{ij} = \cos\theta_{ij}\,, s_{ij} = \sin\theta_{ij}$.
Using the relation 
\begin{align}
  \tan\beta = \frac{v_2}{v_1}\;,
\end{align}
we can perform a rotation to the 'Higgs basis'  (see \cite{Branco:2011iw} and references therein) where only one doublet obtains a vev, 
\begin{align}
  \begin{pmatrix} H_{SM} \\ \tilde{H} \end{pmatrix} = U_\beta \begin{pmatrix} H_1 \\H_2 \end{pmatrix} \,, \quad U_\beta = \begin{pmatrix} \cos\beta & \sin\beta \\ -\sin\beta & \cos\beta \end{pmatrix}\,. \label{eq:rot_higgs}
\end{align}
Here, 
\begin{align}
  H_{SM} = \begin{pmatrix} H_{SM}^+ \\ \frac{1}{\sqrt 2}(v_{SM}+H_{SM}^0) \end{pmatrix}\,, \qquad
  \tilde H = \begin{pmatrix} \tilde H^+ \\ \frac{1}{\sqrt 2}\tilde H^0 \end{pmatrix}\,,
\end{align}
and consequently, $H_{SM}^0$ behaves as the the SM Higgs boson.
Moreover, the Goldstone bosons are given by 
\begin{align}
   G^+ =& \cos\beta H_1^+ + \sin\beta H_2^+\,, \\
   G^0 =& \cos\beta \rho_1 + \sin\beta\rho_2
 \end{align}
 and the charged and CP odd states are
 \begin{align}
  H^+ &= -\sin\beta H_1^+ + \cos\beta H_2^+ \,,\\
  A &= -\sin\beta \rho_1 +\cos\beta \rho_2 \,.
 \end{align}
Combining \eqref{eq:massbasis} and \eqref{eq:rot_higgs}, we find
\begin{align}
  \begin{pmatrix} H_{SM}^0\\\tilde H^0\\S\end{pmatrix} = \tilde U_\beta U^T \begin{pmatrix} h^0\\H^0\\S^0 \end{pmatrix} \,, \quad \tilde U_\beta = \begin{pmatrix} \cos\beta & \sin\beta & 0\\ -\sin\beta & \cos\beta & 0 \\0 & 0 & 1\end{pmatrix}\, \label{eq:higgstomass}
\end{align}
and therefore 
\begin{align}
  H^0_{SM} = \cos\theta_{13}\cos(\beta-\theta_{12})h^0 -\cos\theta_{13}\sin(\beta-\theta_{12}) H^0 + \sin\theta_{13} S^0\,.\label{eq:higgstomass2}
\end{align}
Next, we impose the alignment limit where either $h^0$ or $H^0$ is approximately aligned with $H_{SM}^0$. We will restrict ourselves to the case where $h^0$ is identified as the SM Higgs boson, i.e.  $m_h^0\approx \SI{125}{\giga\electronvolt}$. From \eqref{eq:higgstomass2}, we conclude that the alignment limit modulo $2\pi$ holds for 
\begin{align} 
  \theta_{13} = 0\,, \quad \theta_{12} = \beta\,. \label{eq:alignment}
\end{align}

\subsubsection{Different Vacua}
The potential in \eqref{eq:potential} has 8 different extrema found via the minimizing conditions 
\begin{align}
  \frac{\partial V}{\partial \phi}\Big|_{\phi = \bigl<\phi\bigr>}=0\,, \quad\phi = \{H_1,H_2,S\}\,,\quad \sqrt{2}\bigl<\phi\bigr>={v_1,v_2,v_S}\,.
\end{align}
Using the abbreviations given in \eqref{eq:abbreviations} and 
\begin{align}
\Lambda_{12} &= \lambda_1\lambda_2-\lambda_{345}^2\,,\\
\epsilon &= \lambda_{345}\mu_2^2-\lambda_1\lambda_2 v^2\,,\\
\begin{split}
\alpha &= \mu _S^2+\eta _2^2 \left(2 \lambda _S \left(\mu _2^4 \left(\lambda _{345}^2+\Lambda _{12}\right)+\Lambda _{12} v^4 \left(\lambda _{345}^2+\Lambda _{12}\right)\right.\right.\\
&\left.\left.-\mu _2^2 \left(\delta _{345} \lambda _{345}^2+\lambda _1 v^2 \left(\lambda _{345} \left(\lambda _{345}+\Lambda _2\right)+\Lambda _{12}\right)\right)\right)-\lambda _2 \Lambda _1^2 \mu _S^4\right) \,,
\end{split}\\
\begin{split}
\beta &= -2 \delta _1 \eta _2^3 \mu _S^2+\lambda _S \left(\lambda _S \left(2 \delta _{345} \lambda _{345} \mu _2^2-\delta _2^2 \lambda _1+\lambda _2 \left(\lambda _{345}^2 v^4-\mu _2^4\right)\right)-\Lambda _2^2 \mu _S^4\right)\\&
+2 \eta _2 \lambda _S \mu _S^2 \left(\delta _2 \lambda _1-2 \lambda _{345} \mu _2^2+\lambda _2 \mu _2^2+\lambda _{345}^2 v^2\right)+\eta _2^2 \left(\left(\lambda _2-\lambda _1\right) \mu _S^4\right.\\
&\left.+\lambda _S \left(\mu _2^4+v^4 \left(\lambda _{345}^2+2 \Lambda _{12}\right)-2 \lambda _1 \mu _2^2 v^2\right)\right)-\eta _2^4 \lambda _1 v^4\,,
\end{split}\
\end{align}
the extrema can be found in Tab. \ref{tab:vacuum}.
\begin{table}
  \centering
  \begin{tabular}{c c c c c}
  \toprule
  {$i$} & $v_1^2$ & $v_2^2$ & $v_S^2$ & $\mathcal E_i$\\
  \midrule 
  $1$ & $0 $&$ -\frac{\kappa }{\tilde{\eta }_2} $&$ \frac{\eta _2 \mu _2^2-\lambda _2 \mu _S^2}{\tilde{\eta }_2} $&$ \frac{-2 \eta _2 \mu _2^2 \mu _S^2+\mu _2^4 \lambda _S+\lambda _2 \mu _S^4}{4 \eta _2^2-4 \lambda _2 \lambda _S} $\\
$2$ & $0 $&$ \frac{\mu _2^2}{\lambda _2} $&$ 0 $&$ -\frac{\mu _2^4}{4 \lambda _2} $\\
$3$ & $\frac{\delta _2 \lambda _S-\frac{\eta _2 \left(\lambda _2 \Lambda _1 \mu _S^2+\eta _2 \epsilon \right)}{\Lambda _{12}}}{\sigma } $&$ \frac{\delta _{345} \left(\eta _2^2 \lambda _1-\Lambda _{12} \lambda _S\right)+\eta _2 \lambda _{345} \Lambda _1 \mu _S^2}{\Lambda _{12} \sigma } $&$ 0 $&$ \frac{\alpha }{4 \Lambda _{12} \sigma ^2} $\\
$4$ & $\frac{v^2 \tilde{\eta }_2+\kappa }{\sigma } $&$ \frac{\lambda _{345} v^2 \lambda _S-\kappa }{\sigma } $&$ \frac{\delta _{345} \eta _2-\Lambda _2 \mu _S^2}{\sigma } $&$ \frac{\beta }{4 \sigma ^2} $\\
$5$ & $0 $&$ 0 $&$ \frac{\mu _S^2}{\lambda _S} $&$ -\frac{\mu _S^4}{4 \lambda _S} $\\
$6$ & $\frac{v^2 \left(\lambda _1 \tilde{\eta }_2+\lambda _{345}^2 \lambda _S\right)+\kappa  \Lambda _1}{\lambda _1 \sigma } $&$ 0 $&$ \frac{\mu _S^2}{\lambda _S} $&$ -\frac{\lambda _S \left(v^2 \left(\lambda _1 \tilde{\eta }_2+\lambda _{345}^2 \lambda _S\right)+\kappa  \Lambda _1\right){}^2+\lambda _1 \sigma ^2 \mu _S^4}{4 \lambda _1 \sigma ^2 \lambda _S} $\\
$7$ & $\frac{v^2 \left(\lambda _1 \tilde{\eta }_2+\lambda _{345}^2 \lambda _S\right)+\kappa  \Lambda _1}{\lambda _1 \sigma } $&$ 0 $&$ 0 $&$ -\frac{\left(v^2 \left(\lambda _1 \tilde{\eta }_2+\lambda _{345}^2 \lambda _S\right)+\kappa  \Lambda _1\right){}^2}{4 \lambda _1 \sigma ^2} $\\
$8$ & $0 $&$ 0 $&$ 0 $&$ 0 $\\
  \bottomrule
\end{tabular}
\caption{Extrema energies $\mathcal E_i$ and the corresponding vevs for the potential \eqref{eq:potential} in the case $\eta_1=0$. }
\label{tab:vacuum}
\end{table}
We will focus on the case where all three scalars obtain a vev, $v_{1,2,S} \neq 0$. Consequently, we demand that extremum $\mathcal E_4$ realizes the vacuum, i.e. $\mathcal E_4 < \mathcal E_{1,2,3,5,6,7,8}$. As will be shown in Sec. \ref{sec:muon}, $v_S\neq 0$ results in a small but nonzero mixing between the muon type fermions and introduces a relation between the Yukawa couplings $y, y^\prime$.

\subsubsection{Vacuum Stability}
The potential \eqref{eq:potential} is subject to stability conditions which ensure that the potential is bounded from below. 
The necessary and sufficient conditions for a potential of the type \eqref{eq:potential} which can be found in \cite{Klimenko:1984qx} are given by
  \begin{align}
    \lambda_{1,2,S}>0\,,\quad
  \sqrt{\lambda_S\lambda_1} + \eta_1 > 0 \,,\quad
  \sqrt{\lambda_S\lambda_2} + \eta_2 > 0 \,,\quad
  \sqrt{\lambda_S\lambda_1} + \lambda_{345} > 0 \,,\quad
  \sqrt{\frac{\lambda_1}{\lambda_2}}\eta_2 + \eta_1 \geq 0 \label{eq:stabvac1}
  \end{align}
if $\lambda_4 \geq |\lambda_5|$ and by 
  \begin{align}
    \lambda_{1,2,S}>0\,,\quad
  \sqrt{\lambda_S\lambda_1} + \eta_1 \geq 0 \,,\quad
  \sqrt{\lambda_S\lambda_2} + \eta_2 > 0 \,,\quad
  \sqrt{\lambda_S\lambda_2} - \eta_2 \geq 0\,, \nonumber\\
  - \eta_2\sqrt{\frac{\lambda_1}{\lambda_2}} -\eta_1 \geq0\,,\quad
  \lambda_S \lambda_{345} - \eta_1\eta_2 + \sqrt{(\eta_1^2-\lambda_S\lambda_1)\eta_2^2-\lambda_S\lambda_2)} \geq 0 \label{eq:stabvac2}
  \end{align}
if $\lambda_4 < |\lambda_5|$.

\section{Muon Sector}
\label{sec:muon}
The model features non-diagonal terms at tree level in the extended muon mass matrix, 
\begin{align}
  \lag_{mass} \propto \begin{pmatrix} \overline{(\mu_R)^C} & \overline{(\mu_L)}  & \overline{(M_R)^C} & \overline{(M_L)} \end{pmatrix}
    \underbrace{\begin{pmatrix} 0 & gv_1 & 0 & yv_2 \\ gv_1 & 0 & y^\prime v_S & 0 \\ 0 & y^\prime v_S & 0 & m_M \\ 
    yv_2 & 0 & m_M & 0 \end{pmatrix}}_{=: K} \begin{pmatrix} \mu_R \\ (\mu_L)^C \\M_R \\ (M_L)^C \end{pmatrix}\,.
\end{align}
Defining a Seesaw-type mass matrix $K$, made of $2\times 2$ blocks, as 
\begin{align}
  K &:= \begin{pmatrix} m_\mu & m_{mix} \\ m_{mix}^T & m_M \end{pmatrix}\,, \qquad \mu^f = \begin{pmatrix} \mu_R \\ (\mu_L)^C \end{pmatrix}\,, \qquad M^f = \begin{pmatrix} M_R \\ (M_L)^C \end{pmatrix} \,,
\end{align}
we block diagonalize $K$ in the limit $m_M \gg m_{mix}$, 
yielding
\begin{align}
  K_1 &= -\frac{m_{mix}m_{mix}^T}{m_\mu + m_M}\,,\qquad K_2 = m_\mu +m_M 
\end{align}
where 
\begin{align}
  m_\mu &\sim gv_1\,, \qquad m_{mix} \sim \sqrt{yv_2y^\prime v_S}\,.
  \label{eq:approx}
\end{align} 
With an orthogonal mixing matrix $V$ that diagonalizes $K$ as
\begin{align}
  V^T K V = \text{diag}(K_1, K_2)
\end{align}
where
\begin{align}
  V := \begin{pmatrix}  V_{\mu\mu} & -V_{\mu M} \\ V_{\mu M} &  V_{M M} \end{pmatrix}= \begin{pmatrix}  \cos\alpha & -\sin\alpha \\ \sin\alpha &  \cos\alpha \end{pmatrix}  \,,
  \label{def_v}
\end{align}
we find the mixing angle in the limit $m_M \gg m_{mix}$,
\begin{align}
 \tan\alpha &\approx -\frac{m_{mix}^T}{m_M} \approx -\frac{m_\mu}{m_{mix}}\,.
\end{align}
We conclude that the relation 
\begin{align}
  m_{mix}^T m_{mix} &\approx m_\mu m_M \approx yv_2y^\prime v_S
\end{align}
must hold, thus yielding 
\begin{align}
  K_1 &\approx m_\mu \qquad K_2 \approx m_M\,.
\end{align}
Moreover, the muon-type mass eigenstates are given by
\begin{align} 
  \begin{pmatrix} \mu^m \\ M^m \end{pmatrix} = V^T  \begin{pmatrix} \mu^f \\ M^f \end{pmatrix}\,.
\end{align} 
We note that the couplings in \eqref{lag} can lead to the possibility for direct production of $M_L$ and $M_R$ via $S^0 \to \overline{\mu_L}+ M_R,\,\overline{M_L}+ M_R$, provided that $m_{S^{0'}} > m_M$ and $m_{S^{0'}} > 2m_M$, respectively.

\subsection{Muon anamolous magnetic moment}
\begin{figure}
  \centering
  \begin{subfigure}{0.32\textwidth}
  \centering
  \begin{tikzpicture}
  \begin{feynman}
  \vertex (a1) {\(\mu^-_R\)};
  \vertex[right=1cm of a1] (a2);
  \vertex[above right=1.4cm of a2] (b1);
  \vertex[below right=1.4cm of b1](a3);
  \vertex[right=1cm of a3] (a5) {\(\mu^-_L\)};
  \vertex[above=1.25cm of a5] (c1);
  \vertex[below=0.5em of a2] {\(y\)};
  \vertex[below=0.5em of a3] {\(y^\prime\)};
  \diagram*{
    (a1) --(a2) 
    -- [quarter left, edge label=\(M_L^-\)] (b1) 
    -- [quarter left, edge label=\(M_R^-\), near end,insertion=0.333] (a3) 
    --[scalar, insertion=0.5,edge label=\(h_2{,}\,S^\prime\)] (a2) , 
    (a3) -- (a5), 
    (b1)--[boson,bend left,edge label=\(\gamma\)] (c1), 
  };
  \end{feynman}
  \end{tikzpicture}
  \end{subfigure}
  \begin{subfigure}{0.32\textwidth}
  \centering
  \begin{tikzpicture}
  \begin{feynman}
  \vertex (a1){\(\mu^-_R\)};
  \vertex[right=1cm of a1] (a2);
  \vertex[below=0.5em of a2] {\(y\)};
  \vertex[above right=1.4cm of a2] (b1);
  \vertex[below right=1.4cm of b1](a3);
  \vertex[below=0.5em of a3] {\(y\)};
  \vertex[right=1.5cm of a3] (a5) {\(\mu^-_L\)};
  \vertex[above=1.7cm of a5] (c1);
  \diagram*{
    (a1) -- (a2) 
    -- [quarter left, edge label=\(M_L^-\)] (b1) 
    -- [quarter left, edge label=\(\overline{M_L^-}\)] (a3) 
    --[scalar, edge label=\(h_2\)] (a2) , 
    (a3) --[insertion=0.5] (a5), 
    (b1)--[boson,bend left,edge label=\(\gamma\)] (c1), 
  };
  \end{feynman}
  \end{tikzpicture}
  \end{subfigure}
  \begin{subfigure}{0.3\textwidth}
  \centering
  \begin{tikzpicture}
  \begin{feynman}
  \vertex (a1){\(\mu^-_R\)};
  \vertex[right=1.5cm of a1] (a2);
  \vertex[below=0.5em of a2] {\(y^\prime\)};
  \vertex[above right=1.4cm of a2] (b1);
  \vertex[below right=1.4cm of b1](a3);
  \vertex[below=0.5em of a3] {\(y^\prime\)};
  \vertex[right=1cm of a3] (a5) {\(\mu^-_L\)};
  \vertex[above=1.7cm of a5] (c1);
  \diagram*{
    (a1) --[insertion=0.5] (a2) 
    -- [quarter left, edge label=\(\overline{M_R^-}\)] (b1) 
    -- [quarter left, edge label=\(M_R^-\)] (a3) 
    --[scalar,edge label=\(S^\prime\)] (a2) , 
    (a3) -- (a5), 
    (b1)--[boson,bend left,edge label=\(\gamma\)] (c1), 
  };
  \end{feynman}
  \end{tikzpicture}
  \end{subfigure}
  \caption{Feynman diagrams contributing to $a_\mu$ in the model with the new muon type charged fermions. }
  \label{fig:amu_diagram}
\end{figure}
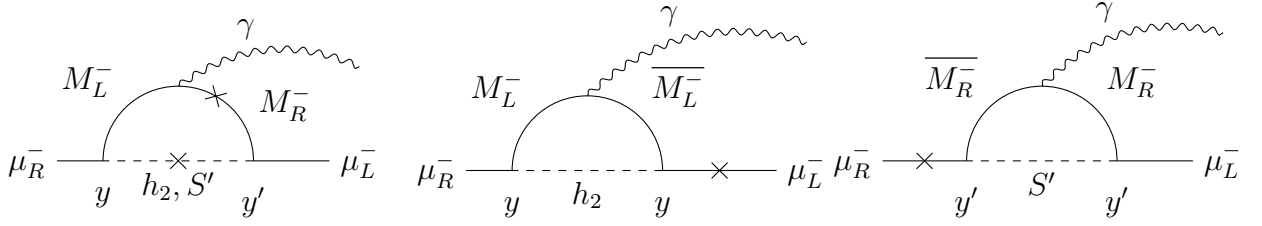

For a general Yukawa interaction between a fermion $F$, a lepton $\ell$ and a scalar $\Phi$ described by the Lagrangian 
\begin{align}
  \lag = \bar F(c_L P_L + c_R P_R) \ell \Phi + \hc\,,
\end{align}
the one-loop contribution to the anomalous magnetic moment of $\ell$ is given by 
\begin{align}
  a_\ell(c_L,c_R,M_F,M_\Phi) = \frac{m_\ell^2}{8 \pi^2} \int_0^1\mathrm dx \frac{\hlf(c_L^2+c_R^2)(x^2-x^3) + \frac{M_F}{m_\ell}c_Lc_R x^2}{m_\ell x^2 + (M_F^2-m_\ell^2)x + M_\Phi^2(1-x)}\,.
\end{align}
In the limit $M_\ell \to 0$, the loop integrals are given by \cite{Athron:2021iuf}
\begin{align}
  I_1(t) &= \frac{t^3 - 6t^2 + 3t + 6t \ln t +2}{3(t-1)^4}\,, \\
  I_2(t) &= \frac{t^2-4t +2 \ln t +3}{(t-1)^3}\,,
\end{align}
resulting in 
\begin{align}
  a_\ell(c_L,c_R,M_F,M_\Phi)  &= \frac{m_\ell^2}{16 \pi^2 M_\Phi^2}\left[  \hlf(c_L^2+c_R^2)I_1\left( \frac{M_F^2}{M_\Phi^2} \right) + \frac{M_F}{m_\ell} c_Lc_R I_2\left( \frac{M_F^2}{M_\Phi^2} \right) \right]\,. \label{eq:amu_gen}
\end{align}
In our model, we aim for contributions to $a_\mu$ from fermion-scalar loops as shown in \ref{fig:amu_diagram}.
After rotating to the mass basis, the relevant couplings $c_{L_i}, c_{R_i}$ are given by 
\begin{align}
  c_{L_{h^0}} &= y^\prime s_{13}          \,\quad&  c_{R_{h^0}} &= ys_{12}c_{13}\,, \\
  c_{L_{H^0}} &= y^\prime s_{23}c_{13}    \,\quad&  c_{R_{H^0}} &= y(c_{12}c_{23}- s_{12}s_{23}s_{13})\,, \\
  c_{L_{S^0}} &= y^\prime c_{23}c_{12}    \,\quad&  c_{R_{S^0}} &= -y(c_{12}s_{23}+s_{12}c_{23}s_{23})\,.
\end{align}
Using 
\begin{align}
  y \approx \frac{m_\mu m_M}{v_2 v_S y^\prime}\,, \label{eq:hhp}
\end{align}
we can write \eqref{eq:amu_gen} as 
\begin{align}
   a_\ell(c_L,c_R,M_F,M_\Phi)  &= \frac{m_\ell^2}{32 \pi^2 M_\Phi^2}\left[\frac{y^{{\prime}^2} \hat c_L^2}{M_\Phi^2} I_1\left(x \right) + x\left(\frac{m_\ell^2}{v_2^2 v_S^2}\frac{\hat c_L^2}{y^{{\prime}^2}} I_1\left(x \right) + \frac{2\hat c_L \hat c_R}{v_2 v_S}I_2\left(x \right) \right)\right]\label{eq:amu_terms}
\end{align}
where 
\begin{align}
  x&= \frac{M_F^2}{M_\Phi^2}\,, \qquad  \hat c_L = \frac{c_L}{y^\prime}\,, \qquad  \hat c_L = \frac{c_L}{y^\prime}\,.
\end{align}
Finally, we sum over all scalar contributions, yielding
\begin{align}
	a_\mu^{model} =& a_\mu(c_{L_{h^0}},c_{R_{h^0}},M_F,m_{h^0}) +a_\mu(c_{L_{H^0}},c_{R_{H^0}},M_F,m_{H^0}) +a_\mu(c_{L_{S^0}},c_{R_{S^0}},M_F,m_{S^0}) \,.\label{eq:amu}
\end{align}

\section{Constraints}
\label{sec:constraints}
\subsection{Higgs Decays}
\label{sec:higgs}
So far, we have assumed that the Higgs boson with $m_{h^0}\approx \SI{125}{\giga\electronvolt}$ behaves just like the SM Higgs
boson. However, introducing new fermions may affect the properties of the $\SI{125}{\giga\electronvolt}$ Higgs and thereby hint at physics beyond the SM. In the following, we examine the effect of the new muon type fermions on the decay channels $h^0 \to\mu\mu, \gamma\gamma, \gamma Z$. In Sec. \ref{sec:htomumu}-\ref{sec:htogz}, we summarize the modification of the decay rates due to the presence of the additional fermions and give the respective experimental limits. For convenience, we define
\begin{align}
  \Delta \Gamma\left(h^0\to a b\right):= \frac{\Gamma\left(h^0\to a b\right)^M}{\Gamma\left(h^0\to a b\right)^{SM}}\,.
\end{align}

\subsubsection{Leptonic Higgs Decays}
\label{sec:htomumu}
In the model with an extended muon sector, leptonic Higgs decays to  SM muons are induced by 
\begin{align}
  \lag \sim \underbrace{\left[ \frac{m_{\mu}}{v_1} V_{\mu\mu}^2 c_{12}c_{13} + yV_{\mu\mu}V_{\mu M} s_{12}c_{13} + y^\prime V_{\mu M}V_{\mu\mu}s_{13} \right]}_{g_{h^0\mu\mu}} h^0 \overline{\mu^m}\mu^m\,,
\end{align}
with a decay rate $\Gamma(h^0\to\mu\mu)^M\sim g_{h^0\mu\mu}$.
Consequently, the decay rate is altered as 
\begin{align}
  \Delta \Gamma(h^0\to\mu\mu) := \frac{\Gamma(h^0\to\mu\mu)^M}{\Gamma(h^0\to\mu\mu)^{SM}} = \left| \frac{v}{m_{\mu}}g_{h^0\mu\mu} \right|^2\,.\label{eq:mumu}
\end{align}
Recent results from CMS \cite{CMS:2020xwi} on Higgs to muon decays at $99.7\%$ CL are given by 
\begin{align}
  \frac{(\sigma \mathcal B(h^0\to \mu\mu))_{obs}}{(\sigma \mathcal B(h^0\to \mu\mu))_{SM}}   = 1.19^{+1.28}_{-1.24} \,. \label{eq:cmshtomumu}
\end{align}

\subsubsection[Higgs decay]{$h^0\to\gamma \gamma$}
\label{sec:htogg}
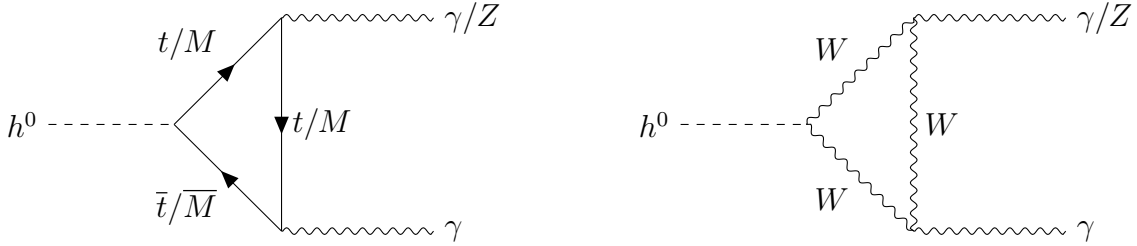
\begin{figure}[H]
\begin{subfigure}{0.49\textwidth}
  \centering
  \begin{tikzpicture}
  \begin{feynman}
  \vertex (a1) {\(h^{0}\)};
  \vertex[right=2cm of a1] (a2);
  \vertex[above right=2cm of a2] (a3);
  \vertex[below right=2cm of a2] (a4); 
  \vertex[right=2cm of a3] (b1){\(\gamma/Z\)};
  \vertex[right=2cm of a4] (b2){\(\gamma\)}; 
  \diagram*{
    (a1) --[scalar] (a2) --[fermion,edge label=\(t/M\)](a3) --[fermion,edge label=\(t/M\)](a4)--[fermion,edge label=\(\overline t/\overline M\)](a2), 
    (a3) -- [boson] (b1), 
    (a4) -- [boson](b2),
  };
  \end{feynman}
  \end{tikzpicture}
\end{subfigure}
\begin{subfigure}{0.49\textwidth}
  \centering
  \begin{tikzpicture}
  \begin{feynman}
  \vertex (a1) {\(h^{0}\)};
  \vertex[right=2cm of a1] (a2);
  \vertex[above right=2cm of a2] (a3);
  \vertex[below right=2cm of a2] (a4); 
  \vertex[right=2cm of a3] (b1){\(\gamma/Z\)};
  \vertex[right=2cm of a4] (b2){\(\gamma\)}; 
  \diagram*{
    (a1) --[scalar] (a2) --[boson,edge label=\(W\)](a3) --[boson,edge label=\(W\)](a4)--[boson,edge label=\(W\)](a2), 
    (a3) -- [boson] (b1), 
    (a4) -- [boson](b2),
  };
  \end{feynman}
  \end{tikzpicture}
  \end{subfigure}
  \caption{Feynman diagrams for $h^0 \to \gamma \gamma(Z)$ at 1-loop order.}
  \label{fig:higgsdiphoton}
\end{figure}

In the SM, the main contribution to the Higgs decay to two photons comes from the diagrams with top-quarks and $W$-bosons propagating in the loop (see Figure \ref{fig:higgsdiphoton}). 
Thus, the decay width in the SM is given by 
\begin{align}
  \Gamma(h^0 \to \gamma\gamma)_{SM} \propto \left| \frac{2}{v} A_1(\tau_W) + \frac{2}{v} N_{c,t} Q_t^2 A_\frac{1}{2}(\tau_t)  \right|^2\,,
\end{align}
where $A_1$ and $A_\frac{1}{2}$ are loop functions, given in \ref{app_loop}, $N_{c,t} = 3$ is the number of color, $Q_t = \frac{2}{3}$ is the top quark electric charge and $\tau_i := 4 \frac{m_i^2}{m_h^2}$. 
Stringent constraints on $h^0\to \gamma\gamma$ come from CMS \cite{CMS:2021kom} where 
\begin{align}
  \frac{(\sigma \mathcal B(h^0\to \gamma\gamma))_{obs}}{(\sigma \mathcal B(h^0\to \gamma\gamma))_{SM}}   = 1.12^{+0.27}_{-0.27}\,\, \label{eq:cmsdiphoton}
\end{align}
at $99.7\%$ CL.
In the model with an extended muon sector, the muon type charged lepton can propagate in the loop, thus potentially changing the Higgs to diphoton rate. 
The relevant terms are given by 
\begin{align}
  \lag \sim \underbrace{\left[ \frac{m_\mu}{v_1}V_{\mu M}^2 c_{12}c_{13}- yV_{MM}V_{\mu M} s_{12}c_{13} - y^\prime V_{\mu M}V_{MM}s_{13}  \right]}_{g_{h^0MM}} h^0 \overline{M^m}M^m\,
\end{align}
and the decay rate reads 
\begin{align}
  \Gamma(h^0 \to \gamma\gamma)_{M} \propto \left| \frac{2}{v} (c_{12}c_{13}c_\beta+s_{12}c_{13}s_\beta) A_1(\tau_W) + \frac{2}{v_1} N_{c,t} Q_t^2 A_\frac{1}{2}(\tau_t)+ \frac{2g_{h^0M M}}{m_M} Q_M^2 A_\frac{1}{2}(\tau_M)  \right|^2\,. \label{eq:gammagamma}
\end{align}

\subsubsection[Higgs decay]{Higgs decay $h^0\to\gamma Z$}
\label{sec:htogz}
Finally, in a similar manner to the diphoton decay, the Higgs boson can also decay to a photon and a $Z$ boson with the main SM contribution coming from the diagrams with top-quarks and $W$-bosons propagating in the loop of \ref{fig:higgsdiphoton}. 
The SM decay width is given by 
\begin{align}
  \Gamma(h^0 \to\gamma Z)_{SM} = \left|\frac{2}{v}\cot\vartheta_WA_1(\tau_W, \lambda_W) + N_{c,t}\frac{2}{v}\frac{2Q_t}{\sin\vartheta_W\cos\vartheta_W}(T_3^{(t)}-2Q_t\sin\vartheta_W^2)A_\frac{1}{2}(\tau_t,\lambda_t)\right|^2
\end{align}
with stringent constraints placed by CMS \cite{CMS:2021wog} at $99.7\%$ CL,  
\begin{align}
  \frac{(\sigma(pp \to h^0) \mathcal B(h^0\to \gamma Z))_{obs}}{(\sigma(pp \to h^0) \mathcal B(h^0\to \gamma Z))_{SM}}   = 2.4^{+2.7}_{-2.7}\,.\label{eq:cmsgammaz}
\end{align}
In our model with an extended muon sector, the $h^0 \to \gamma Z$-rate is given by 
\begin{align}
  \Gamma(h^0 \to\gamma Z)_{SM} &= \Big|\frac{2}{v}\cot\vartheta_W (c_{12}c_{13}c_\beta+s_{12}c_{13}s_\beta) A_1(\tau_W, \lambda_W) \nonumber\\ &+ N_{c,t}\frac{2}{v_1}\frac{2Q_t}{\sin\vartheta_W\cos\vartheta_W}(T_3^{(t)}-2Q_t\sin\vartheta_W^2)A_\frac{1}{2}(\tau_t,\lambda_t) \nonumber\\
  &+\frac{2g_{h^0M M}}{m_M}\frac{2Q_M}{\sin\vartheta_W\cos\vartheta_W}(2T_3^{(M)}-2Q_M\sin\vartheta_W^2)A_\frac{1}{2}(\tau_M,\lambda_M)
  \Big|^2\,. \label{eq:gammaz}
\end{align}

\subsection{Lepton Flavor Violation}
\begin{figure}[h]
  \centering
  \begin{tikzpicture}
  \begin{feynman}
  \vertex (a1) {\(\ell^\prime\)};
  \vertex[right=2cm of a1] (a2);
  \vertex[above right=2cm of a2] (b1);
  \vertex[below right=2cm of b1](a3);
  \vertex[right=2cm of a3] (a5) {\(\ell\)};
  \vertex[above=2.5cm of a5] (c1);
  \diagram*{
    (a1) -- [fermion](a2) 
    -- [boson, quarter left, edge label=\(W\)] (b1) 
    -- [boson, quarter left] (a3) 
    --[anti fermion,edge label=\(\nu_i\)] (a2) , 
    (a3) -- [fermion] (a5), 
    (b1)--[boson,edge label=\(\gamma\)] (c1), 
  };
  \end{feynman}
  \end{tikzpicture}
  \caption{Feynman diagram for $\ell^\prime \to \gamma\ell$ at 1-loop-order where $\ell^\prime$, $\ell$ are charged SM leptons with one of them being a muon and $\nu_i$ is a neutrino in its mass eigenstate.}
  \label{fig:radiativeleptondecay}
\end{figure}
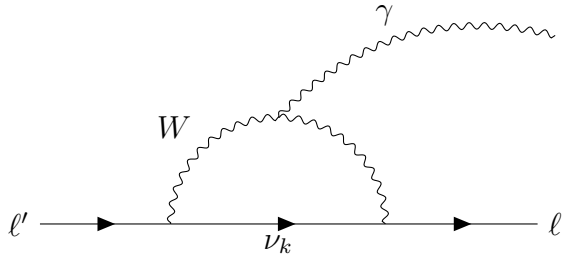

\begin{table}
\centering
\begin{tabular}{c c c}
\toprule
  {Process} & Limit & Reference \\
  \midrule 
  $\mu\to e\gamma$ & $\num{4.2 e-13}$ & \cite{MEG:2016leq} \\
  $\tau \to e\gamma$ & $\num{3.3 e-8}$ & \cite{BaBar:2009hkt} \\
  $\tau \to \mu\gamma$ & $\num{4.4 e-8}$ & \cite{BaBar:2009hkt} \\
  $\mu \to 3e$ & $\num{1.0 e-12}$ & \cite{BERTL19851} \\
  $\tau \to 3e$ & $\num{2.7e-8}$ & \cite{Hayasaka:2010np} \\
  $\tau \to 3\mu$& $\num{2.1e-8}$ & \cite{Hayasaka:2010np} \\
  $\mu \mathrm{Au} \to e \mathrm{Au}$ & $\num{7.0 e-13}$ & \cite{SINDRUMII:2006dvw} \\
  $\mu \mathrm{Ti} \to e \mathrm{Ti}$ & $\num{6.1 e-13}$ & \cite{Wintz:1998rp} \\
  \bottomrule
\end{tabular}
\caption{Overview of current limits on LFV decays and capture rates (bottom two rows). A more extensive list of LFV violating processes can be found in \cite{ParticleDataGroup:2024cfk}.}
\label{tab:lfv}
\end{table}
In general, the considered model with an extended muon sector can give contributions to lepton flavor violating (LFV) processes such as charged LFV decays $\ell \to \ell\prime\gamma$, charged three body decays $\ell \to \ell_1\bar\ell_2\ell_3$ and $\mu\to e$ conversion via nuclei, provided that the model is extended in a suitable way to account for the observed neutrino masses. Current experimental bounds can be found in Tab. \ref{tab:lfv} .
The details depend strongly on the mixing between mass and flavor eigenstates in the neutrino sector and therefore on the underlying mechanism that generates the neutrino mass. As a dedicated analysis is beyond the scope of this paper, we restrict ourselves to a qualitative discussion based on the assumption that some unknown mechanism generates neutrino masses.\footnote{We stress that depending on the mechanism employed, the conclusions can drastically change.} We start the discussion by reviewing the situation in the Standard Model extended with neutrino masses and compare it to the model with an extended muon sector.
As the flavor eigenstates of the neutrinos differ from the mass eigenstates, neutrino oscillations occur and the flavor eigenstates can be written as a sum over the neutrino mass eigenstates as 
\begin{align}
   \ket{\nu_\alpha} = \sum_k U_{\alpha k}\ket{\nu_k}
\end{align}
where $\alpha = {e,\mu,\tau}$, $k = 1,2,3$ in the SM extended by neutrino masses ($\nu\mathrm{SM}$), and $\alpha = {e,\mu,\tau, M^L, M^R}$, $k = 1,2,3,4,5$ in the model with an extended muon sector. As can be seen in Fig. \ref{fig:radiativeleptondecay}, the mixing induces LFV decays $\ell \to \ell^\prime\gamma$ at 1-loop-order.\footnote{Note that there is no contribution from a LFV decay with $M_L, M_R, h^0$, $H^0$ and $S^0$ propagating in the loop similar to Fig. \ref{fig:amu_diagram} as $e$ and $\tau$ do not mix with the new muon type charged leptons.} For simplicity, we focus on $\mu \to e\gamma$ where $\mu$ is a SM muon in its mass eigenstate. The decay width is proportional to a combination of loop functions and $U,V$ matrix elements as \cite{Cheng:1980tp, Minkowski:1977sc}
\begin{align}
	\mathrm{Br}(\mu \to e\gamma) &\propto |G_{\mu e}|^2\,, \label{eq:mutoegamma}\\
	G_{\mu e} &= \sum_i\sum_\beta^{\mu, M} U_{ei}U_{\beta i}^*V_{\mu\beta} G_\gamma(x_i)\,,
\end{align}
where $x_i = \frac{m_i^2}{m_W^2}$ and $m_i$ are the neutrino masses while $G_\gamma(x_i)$ is a loop function, 
\begin{align} 
	G_\gamma(x_i) = - \frac{x(2x^2 + 5x-1)}{4(1-x)^3} - \frac{3x^3}{2(1-x)^4}	\ln x\,.
\end{align}
In the asymptotic limit, we can express $G_\gamma(x_i)$ as 
\begin{align}
	G_\gamma(x) \approx  
	\begin{cases}
		\frac{x}{4}\,,\quad &x\ll1 \\
		\frac{1}{2}\,,\quad &x\gg1 \,.
	\end{cases}
\end{align}
In the $\nu\mathrm{SM}$, the sum over $\beta$ in \eqref{eq:mutoegamma} vanishes and with $i = 1,2,3$, the $\mu \to e\gamma$ rate is strongly suppressed via the GIM mechanism, 
\begin{align}
	\mathrm{Br}(\mu \to e\gamma)_{\nu\mathrm{SM}} &\propto \left|\sum_{i = 2}^3 \frac{U_{ei}U_{\mu i}^*\Delta m_{i1}^2}{m_W^2}  \right|\,,
\end{align}
where $\Delta m_{i1}^2:= m_i^2-m_1^2$ are the neutrino mass-squared differences. The resulting numerical value of the branching ratio is $\mathrm{Br}(\mu \to e\gamma)_{\nu\mathrm{SM}}\sim \num{e-54}$ and thus practically unobservable. 
On the other hand, in the model discussed here we find 
\begin{align}
	G_{\mu e} = \sum_{i = 2}^3 \frac{\Delta m_{i1}^2}{4 m_W^2} U_{ei} \left(V_{\mu\mu} U_{\mu i}^* + V_{\mu M} U_{Mi}^* \right) + \sum_{i = 4}^5 U_{ei} \left[ G(x_i) - \frac{m_1^2}{4m_W^2} \right]\left( V_{\mu\mu}U_{\mu i}^* + V_{\mu M} U_{Mi}^* \right)\,. \label{eq:gmu:model}
\end{align}
In the first term in \eqref{eq:gmu:model}, the GIM suppression is prevalent, irrespective of the neutrino mixing, and it is therefore of order $\mathcal{O}(\sfrac{\Delta m_{i1}^2}{m_W^2})$, similarly to the $\nu\mathrm{SM}$. The second term is more evolved as it strongly depends on whether $m_{4,5}$ are above or below the electroweak scale, the mixing between $\nu_{e,\mu,\tau}$ and $\nu_{M_{L,R}}$ and $V_{\mu\beta}$. While the former two quantities depend on the details of the neutrino sector, the latter is determined already from the charged sector with $V_{\mu\mu} \gg V_{\mu M}$. If we assume that mixing between the SM neutrinos and $\nu_{M_{L,R}}$ is small so that $U_{\mu i} \ll U_{M i}, i=4,5$, it is apparent that this suppresses the term proportional to $V_{\mu\mu}$ while $V_{\mu M}U_{Mi}^*$ is suppressed by the smalls mixing in the muon sector. Hence, even if $m_{4,5}$ are above the electroweak scale and the GIM mechanism unoperative, we generally expect that the branching ratio is way below current senitivity limits, similar to LFV in the type I Seesaw mechanism (see e.g. \cite{Ilakovac:1994kj, Alonso:2012ji}).\\
The rate of $\mu \to e$ conversion is much more complicated than $\mu \to e \gamma$ and depends significantly on the nucleus and the atomic mass and number, $A$ and $Z$, respectively. However, as can be seen in Fig. \ref{fig:mutoeconversion}, the conversion rate depends on the same mixing matrices as $\mu \to e \gamma$. 
Provided that $m_{4,5}$ are above the electroweak scale, we therefore expect a contribution to the capture rate that depends similarly on $U$ and $V$ as the second term in \eqref{eq:gmu:model} and thus the same conclusion holds. \\
Finally, let us comment on the three-body decays $\ell \to \ell_1\bar\ell_2\ell_3$. The amplitudes receive contributions from several diagrams, see \cite{Ilakovac:1994kj} for details, and is highly non-trivial. It is however immediatly clear that each muon in the initial or final state gives rise to a sum over $\sum_{\beta = \mu,M} V_{\mu\beta}$ while the neutrino propagators introduce dependencies on $U$. Naively, we therefore expect that similar arguments regarding the contributions from the extended muon sector as above hold but stress that the details are significantly more complicated and would require a detailed analysis of the neutrino sector.  
\begin{figure}
  \centering
  \begin{subfigure}{0.49\textwidth}
  \centering
  \begin{tikzpicture}
  \begin{feynman}
  \vertex (a1) {\(\mu\)};
  \vertex[right=2cm of a1] (a2);
  \vertex[right=2cm of a2] (a3);
  \vertex[right=1cm of a2] (i1);
  \vertex[below=1cm of i1] (c1);
  \vertex[right=2cm of a3] (a4){\(e\)};
  \vertex[below=1cm of c1] (b2);
  \vertex[left=3cm of b2] (b1){\(u/d\)};
  \vertex[right=3cm of b2] (b3){\(u/d\)};
  \diagram*{
    (a1) -- [fermion](a2) -- [fermion, edge label=\(\nu_i\)](a3) -- [fermion](a4),
    (b1) -- [fermion](b2) -- [fermion](b3),
    (a2)   -- [boson, quarter right] (c1),
    (a3)   -- [boson, quarter left, edge label=\(W\)] (c1), 
    (c1) --[boson,edge label=\(\gamma/Z\)] (b2), 
  };
  \end{feynman}
  \end{tikzpicture}
  \end{subfigure}
  \begin{subfigure}{0.49\textwidth}
  \centering
  \begin{tikzpicture}
  \begin{feynman}
  \vertex (a1) {\(\mu\)};
  \vertex[right=2cm of a1] (a2);
  \vertex[right=2cm of a2] (a3);
  \vertex[right=1cm of a2] (i1);
  \vertex[below=1cm of i1] (c1);
  \vertex[right=2cm of a3] (a4){\(e\)};
  \vertex[below=1cm of c1] (b2);
  \vertex[left=3cm of b2] (b1){\(u/d\)};
  \vertex[right=3cm of b2] (b3){\(u/d\)};
  \diagram*{
    (a1) -- [fermion](a2) -- [boson, edge label=\(W\)](a3) -- [fermion](a4),
    (b1) -- [fermion](b2) -- [fermion](b3),
    (a2)   -- [fermion, quarter right, edge label'=\(\nu_i\)] (c1),
    (c1)   -- [fermion, quarter right, edge label'=\(\nu_j\)] (a3), 
    (c1) --[boson,edge label=\(\gamma/Z\)] (b2), 
  };
  \end{feynman}
  \end{tikzpicture}
  \end{subfigure}\\
    \centering
  \begin{subfigure}{0.49\textwidth}
  \centering
  \begin{tikzpicture}
  \begin{feynman}
  \vertex (a1) {\(\mu\)};
  \vertex[right=2cm of a1] (a2);
  \vertex[right=2cm of a2] (a3);
  \vertex[right=2cm of a3] (a4){\(e\)};
  \vertex[below=2cm of a1] (b1){\(u\)};
  \vertex[below=2cm of a2] (b2);
  \vertex[below=2cm of a3] (b3);
  \vertex[below=2cm of a4] (b4){\(u\)};
  \diagram*{
    (a1) -- [fermion](a2) -- [fermion, edge label=\(\nu_i\)](a3) -- [fermion](a4),
    (b1) -- [fermion](b2) -- [fermion, edge label'=\(d\)](b3) -- [fermion](b4),
    (a2) -- [boson, edge label=\(W\)](b2), 
    (a3) -- [boson, edge label=\(W\)](b3)
  };
  \end{feynman}
  \end{tikzpicture}
  \end{subfigure}
  \begin{subfigure}{0.49\textwidth}
  \centering
  \begin{tikzpicture}
  \begin{feynman}
  \vertex (a1) {\(\mu\)};
  \vertex[right=2cm of a1] (a2);
  \vertex[right=2cm of a2] (a3);
  \vertex[right=2cm of a3] (a4){\(e\)};
  \vertex[below=2cm of a1] (b1){\(d\)};
  \vertex[below=2cm of a2] (b2);
  \vertex[below=2cm of a3] (b3);
  \vertex[below=2cm of a4] (b4){\(d\)};
  \diagram*{
    (a1) -- [fermion](a2) -- [fermion, edge label=\(\nu_i\)](a3) -- [fermion](a4),
    (b1) -- [fermion](b2) -- [fermion, edge label'=\(u\)](b3) -- [fermion](b4),
    (a2) -- [boson, edge label'=\(W\), near start](b3), 
    (a3) -- [boson, edge label=\(W\), near start](b2)
  };
  \end{feynman}
  \end{tikzpicture}
  \end{subfigure}
  \caption{Feynman diagrams for the $\mu\to e$ conversion where $\nu_i$ is a neutrino in its mass eigenstate. }
  \label{fig:mutoeconversion}
\end{figure}

\subsection{Collider Signals}
\label{sec:collider}
Finally, let us shortly comment on possible collider signals. At $pp$ colliders, vectorlike leptons can be produced via $\gamma, Z$ or $W$ exchange and decay via gauge interactions, $M \to W \nu$, resulting in multi-lepton final states with missing energy as for example shown in Fig. \ref{fig:collider}. To date, direct collider searches are focused on the search for vectorlike $\tau$ leptons \cite{ATLAS:2023sbu,Kumar:2015tna,CMS:2022cpe,CMS:2019hsm}, with the most stringent bound excluding masses up to $\SI{1040}{\giga\electronvolt}$\cite{CMS:2022nty}. Significantly weaker bounds that are applicable to the model discussed here are searches for additional heavy leptons at LEP which place a lower bound on the mass at around $\SI{100}{\giga\electronvolt}$ \cite{L3:2001xsz}. 
Results from LHC Run-2 can be used to project the sensitivity to the HL-LHC for models including vectorlike muons. With a total integrated luminosity of $\SI{3000}{\per\femto\barn}$ at $\sqrt{s} = \SI{14}{\tera\electronvolt}$, CMS expects to exlude vectorlike muons at $\SI{95}{\percent}$ CL up to a mass of $\SI{1630}{\giga\electronvolt}$ \cite{CMS:2024bni}. 
\begin{figure}
  \centering
  \begin{subfigure}{0.4\textwidth}
  \centering
  \begin{tikzpicture}
  \begin{feynman}
  \vertex (b1) ;
  \vertex[above left=1.5cm of b1] (a1){\(d\)};
  \vertex[below left=1.5cm of b1] (c1){\(\overline u\)};
  \vertex[right=1.5cm of b1] (b2);
  \vertex[above right=1.5cm of b2] (a2);
  \vertex[below right=1.5cm of b2] (c2){\(\nu\)};
  \vertex[above right=1.5cm of a2] (d1){\(\nu\)};
  \vertex[right=1.5cm of a2] (a3);
  \vertex[above right=0.75cm of a3] (f1){\(\ell\)};
  \vertex[below right=0.75cm of a3] (f2){\(\overline\nu\)};
  \diagram*{
    (a1) -- [fermion](b1) --[fermion](c1), 
    (b1) --[boson, edge label=\(W\)](b2), 
	(c2) -- [fermion](b2) -- [fermion, edge label=\(M\)](a2) --[fermion](d1),
    (a2) -- [boson, edge label'=\(W\)](a3), 
    (f2) -- [fermion](a3) -- [fermion](f1),

  };
  \end{feynman}
  \end{tikzpicture}
  \end{subfigure}\hspace{0.4cm}
  \begin{subfigure}{0.4\textwidth}
  \centering
  \begin{tikzpicture}
  \begin{feynman}
  \vertex (b1) ;
  \vertex[above left=1.5cm of b1] (a1){\(q\)};
  \vertex[below left=1.5cm of b1] (c1){\(\overline q\)};
  \vertex[right=1.5cm of b1] (b2);
  \vertex[above right=1.5cm of b2] (a2);
  \vertex[below right=1.5cm of b2] (c2);
  \vertex[above right=1.5cm of a2] (d1){\(\nu\)};
  \vertex[right=1.5cm of a2] (a3);
  \vertex[above right=0.75cm of a3] (f1){\(\ell\)};
  \vertex[below right=0.75cm of a3] (f2){\(\overline\nu\)};
  \vertex[right=1.5cm of c2] (c3);
  \vertex[above right=0.75cm of c3] (g1){\(\ell\)};
  \vertex[below right=0.75cm of c3] (g2){\(\overline\nu\)};
  \vertex[below right=1.5cm of c2] (e1){\(\nu\)};
  \diagram*{
    (a1) -- [fermion](b1) --[fermion](c1), 
    (b1) --[boson, edge label=\(\gamma/Z\)](b2), 
    (e1) -- [fermion](c2) -- [fermion, edge label=\(\overline M\)](b2) -- [fermion, edge label=\(M\)](a2) --[fermion](d1),
    (a2) -- [boson, edge label'=\(W\)](a3), 
    (c2) -- [boson, edge label=\(W\)](c3), 
    (f2) -- [fermion](a3) -- [fermion](f1),
    (g2) -- [fermion](c3) -- [fermion](g1),
  };
  \end{feynman}
  \end{tikzpicture}
  \end{subfigure}
  \caption{Examples of signal channel feynman diagrams with $\ell + E_{miss}$ (left) and $2\ell + E_{miss}$ (right) signatures.}
  \label{fig:collider}
\end{figure}

\section{Results}
\label{sec:results}
\begin{figure}[h]
  \centering
    \includegraphics[width=0.9\textwidth]{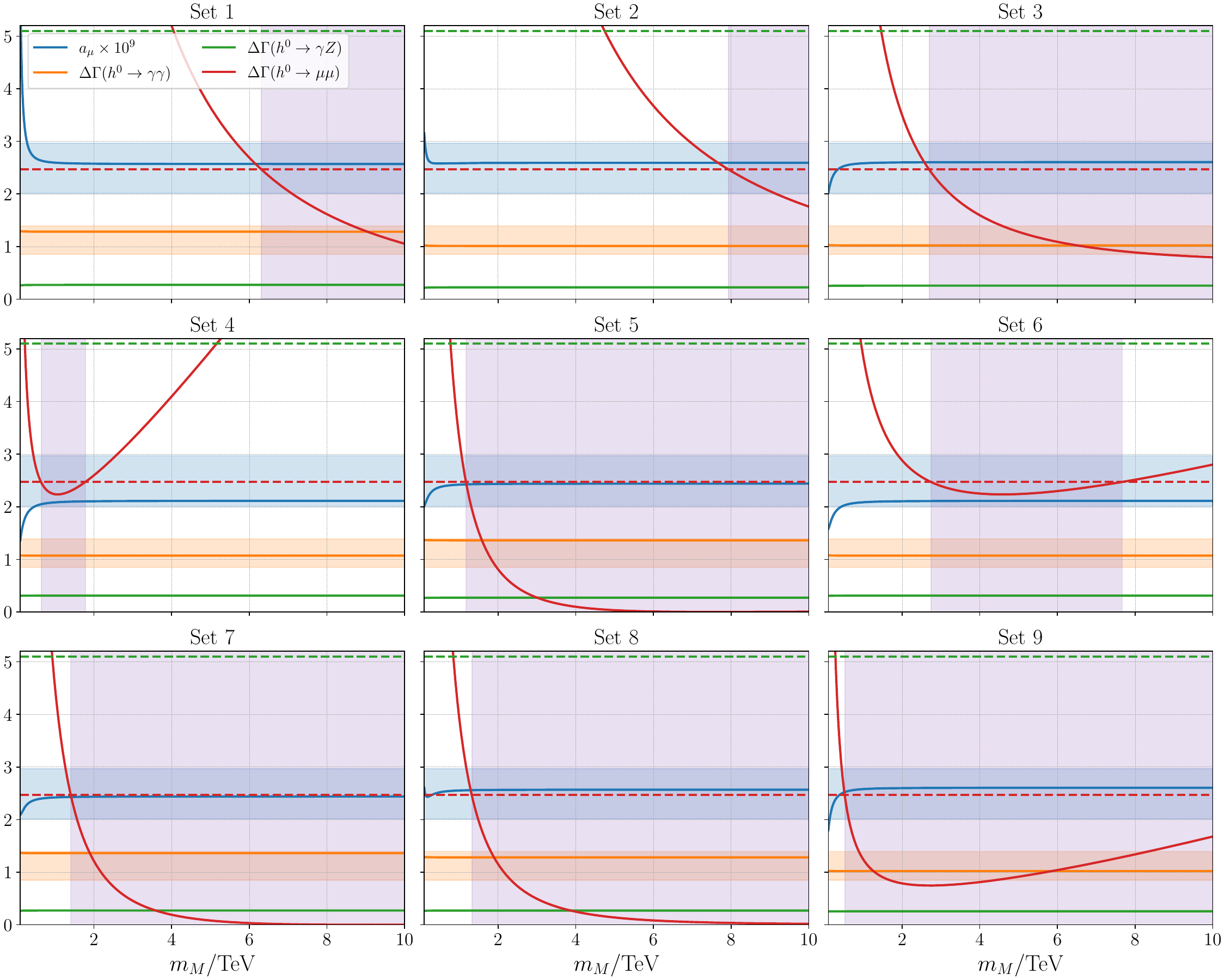}
  \caption{Muon anomalous magnetic moment $a_\mu$ as a function of the vectorlike lepton mass $m_M$ for nine exemplary data sets. In the purple region, $a_\mu$ can be explained and constraints from Higgs decays are fulfilled. The blue region corresponds to the error interval of $\Delta a_\mu$. The orange region corresponds to the limits on $h^0 \to \gamma\gamma$. The red (green) dashed line indicates the upper limit on $h^0 \to \gamma Z$($h^0 \to \mu\mu$).}
  \label{fig:constraintshiggs}
\end{figure}

In the following we perform random scans of the parameter space to identify consistent sets of parameters that explain the anomalous magnetic moment
of the muon and are compatible with existing experimental contraints. To do so, we first determine a consistent set of parameters in the scalar sector.
Next, we scan for values of the vectorlike fermion mass $m_M$ and couplings $y$, $y'$ that combined with the set of parameters identified as consistent above produce an anomalous magnetic moment of the muon in the interval allowed by the experimental result \eqref{eq:amuexp} while being compatible with the experimental constraints from Higgs decays discussed in Sec. \ref{sec:htomumu}-\ref{sec:htogz}. \\
The free parameters in the scalar sector are varied within the ranges
\begin{align}
	0 \leq \lambda_{1,2,S} \leq 1\,,\quad -1\leq \lambda_{3,4,5}\leq 1\,, \quad -1\leq\eta_2\leq1\,, \nonumber\\\quad \eta_1 = 0\,, \quad \frac{v_{SM}^2}{1+50^2} \leq v_1^2 \leq \frac{v_{SM}^2}{2}\,,\quad 0\leq\mu_S\leq \SI{100}{\giga\electronvolt}\,,
\end{align}
demanding that the constraints from vacuum stability, \eqref{eq:stabvac1} and \eqref{eq:stabvac2}, are fulfilled, that extremum $\mathcal E_4$ from Tab. \ref{tab:vacuum} realizes the vacuum and that $m_{h^0}, m_{H^0}, m_{S^0}>0$. Moreover, we used $1\leq \tan\beta \leq 50$ as suggested by LHC searches.
In order to maintain the alignment limit \eqref{eq:alignment}, we impose the limits
\begin{align}
	0.9\leq \cos\theta_{13} \leq 1\,, \qquad \cos\beta-0.1\leq \cos\theta_{12} \leq \cos\beta+0.1\,.  \label{eq:alignment_random}
\end{align}
We find that a wide range of parameters is possible within the previously mentioned constraints. Generally, we observe that values of $0 \lesssim v_S  \lesssim \SI{200}{\giga\electronvolt}$, $0 \lesssim m_{S^0} \lesssim \SI{300}{\giga\electronvolt}$ and $1  \lesssim \tan\beta  \lesssim 20$ are favored, although outliers exist where $300 \lesssim v_S  \lesssim \SI{1400}{\giga\electronvolt}$ and $20  \lesssim \tan\beta  \lesssim 50$.
\begin{figure}[h]
\centering
  \includegraphics[width=\textwidth]{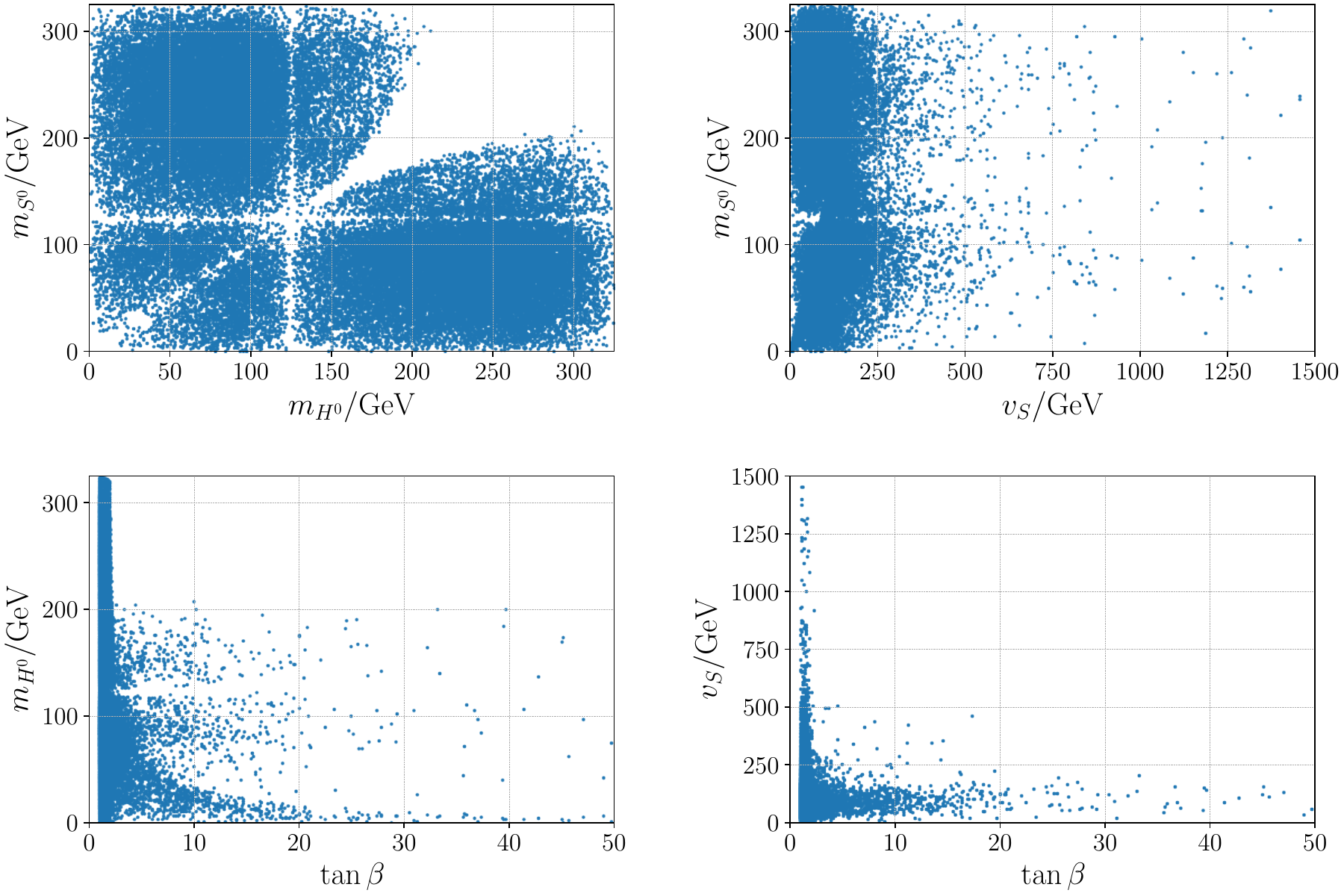}
  \caption{Regions allowed by the constraints from Sec. \ref{sec:model} with an alignment limit according to \eqref{eq:alignment_random}.}
  \label{fig:scatterscalar}
\end{figure}
Using \eqref{eq:amu} and the data sets that are compatible with the scalar sector presented in Fig. \ref{fig:scatterscalar}, we perform a scan over $m_M$ in the range $\SI{100}{\giga\electronvolt}\leq m_M \leq \SI{10}{\tera\electronvolt}$ \cite{L3:2001xsz} and determine $y^\prime$ so that $a_\mu^{model} = \Delta a_\mu$ is satisfied\footnote{Note that $y$ is determined via the relation given in \eqref{eq:hhp}} and constraints from Higgs decays are met.\footnote{Note that for $\Delta \Gamma\left(h^0 \to \mu\mu\right)$ and $\Delta \Gamma\left(h^0 \to \gamma Z\right)$ we impose $\Delta \Gamma\left(h^0 \to \gamma Z, \mu\mu\right) = 0$ as a lower bound.} In order to avoid conflicts with perturbativity we restrict the Yukawa couplings to 
 \begin{align}
 -3\lesssim y, y^\prime \lesssim 3\,.
\end{align}
In Fig. \ref{fig:constraintshiggs}, we show $a_\mu^{model}(m_M)$ for 9 exemplary chosen data sets\footnote{The corresponding data sets are given in Tab. \ref{tab:amupars} in App. \ref{app:data}}.
It is evident that the presented model can give a viable contribution to the muon anamolous magnetic moment while being compatible with constraints from Higgs decays over a large range of $m_M$. 
However, we also note that the respective constraints depend not only on $m_M$ but also on the specific combination of parameters in section\eqref{sec:constraints}  that enter  equations  \eqref{eq:mumu}, \eqref{eq:gammagamma} and \eqref{eq:gammaz}, respectively.
As discussed in Sec. \ref{sec:collider}, future searches for vectorlike muons at CMS \cite{CMS:2024bni} could exclude masses up to $M_M \sim \SI{1.6}{\tera\electronvolt}$. 
This bound would exclude some parameter sets, for example set 4 in Fig. \ref{fig:constraintshiggs}, or not affect the viability of certain parameter sets at all, e.g. set 6 in Fig. \ref{fig:constraintshiggs}.

\section{Conclusion}
\label{sec:conclusion}
In this paper we have investigated a Standard Model extension with a vector-like lepton in the ${\cal O}(10~{\rm TeV})$ region, a scalar singlet and an additional scalar doublet. The scenario can be motivated as the infrared limit of $E_6$ and Trinification models and provides an elegant and natural explanation of the recently announced measurement of the muon anomalous magnetic moment. 
We have carefully studied the model's parameter space with regard to vacuum stability and compatibility with constraints from various Higgs decays. We find that the contribution to the muon anomalous magnetic moment and the contributions to the Higgs decay rates depend strongly on the full set of parameters. 
While $h^0 \to \mu\mu$ can be either enhanced or suppressed, we find that $h^0\to \gamma Z$ can be strongly suppressed while $h^0 \to \gamma\gamma$ tends to be slightly enhanced. Nevertheless, we find various parameter sets that can explain the muon anomalous magnetic moment while being in agreement with constraints from Higgs decays. 
Moreover, we do not find large contributions to lepton flavor non-universality or
charged lepton flavor violating radiative decays such as $\mu \rightarrow e \gamma$ or $\tau \rightarrow \mu \gamma$. We conclude that it will be interesting to search for the direct production of new scalars,  vectorlike leptons and deviations from the SM predictions of Higgs decays at the LHC and at future collider experiments. 

\section*{Acknowledgements}
The work of TWK was supported by US DOE grant DE-SC0019235.
The work of TB was supported by the \textit{Studienstiftung des deutschen Volkes}.

\appendix
\section{Appendix}
\subsection{Higgs Decay Rates }
Defining $\tau_i = 4\frac{m_i^2}{m_h}^2, i=V,f$, the diphoton decay width for a general model including spin$-1$ particles $V$ and fermions $f$ with electric charge $Q_i$ coupling to the Higgs is given by \cite{Carena_2012} 
\begin{align}
  \Gamma(h^0 \to \gamma\gamma) \propto \left| \frac{g_{hVV}}{m_V^2}Q_V^2 A_1(\tau_V) + \frac{2g_{hf\bar f}}{m_f} N_{c,f} Q_f^2 A_\frac{1}{2}(\tau_f)  \right|^2\;.
\end{align}
The $h^0\to\gamma Z$ decay width for a general model including spin$-1$ particles $V$ and fermions $f$ coupling to the Higgs is given by \cite{Carena_2012} 
\begin{align}
\Gamma(h^0 \to\gamma Z) = \left|\frac{g_{hVV}}{m_V^2}g_{ZVV}A_1(\tau_V, \lambda_V) + N_{c,f}\frac{2g_{hf\bar f}}{m_f}(2Q_f)(g_{Zll}^f + g_{Zrr}^f)A_\frac{1}{2}(\tau_f,\lambda_f)\right|^2
\end{align}
where $\lambda_i = 4\frac{m_i^2}{m_Z^2}$ and 
\begin{align}
  g_{Zkk}^f = \frac{1}{\sin\vartheta_W\cos\vartheta_W}(T_3^{(k,f)} - Q_f\sin\vartheta_W^2)\, \qquad k = l,r
\end{align}
are the couplings of the $Z$ to left- and right-handed particles with weak isospin $T_3^{k,f}$, respectively.
\subsection{Loop Functions}
\label{app_loop}
The loop functions used in section \ref{sec:higgs} are given by \cite{Gunion:1989we}
\begin{align}
  A_1(x) &= -x^2\left( 2x^{-2} + 3x^{-1} + 3(2x^{-1}-1)f(x^{-1}) \right) \\
  A_{\frac{1}{2}}(x) &= 2x^2 \left( x^{-1} + \left( x^{-1} - 1 \right)f(x^{-1}) \right) \\
  A_1(x,y) &= 4(3-\tan^2\vartheta_W)I_2(x,y) + [(1-\frac{2}{x})\tan^2\vartheta_W - (5+\frac{2}{x}]I_1(x,y)\\
  A_\frac{1}{2}(x,y) &= I_1(x,y)-I_2(x,y)\\
  I_1(x,y) &= \frac{xy}{2(x-y)} + \frac{x^2y^2}{2(x-y)^2}[f(x^{-1})  - f(y^{-1})] + \frac{x^2y}{(x-y)^2}[g(x^{-1})  - g(y^{-1})] \\
  I_2(x,y) &= - \frac{xy}{2(x-y)}[f(x^{-1})  - f(y^{-1})] \\
  f(x) &= \arcsin^2\sqrt x \\
  g(x) &= \sqrt{x^{-1} - 1} \arcsin\sqrt x
\end{align}

\subsection{Exemplary Data Sets}
\label{app:data}
\begin{landscape}
\begin{table}
\begin{tabular}{c c c c c c c c c c}
\toprule
  {Parameter} & $1$& $2$ & $3$ & $4$ & $5$ & $6$& $7$ & $8$& $9$  \\
  \midrule 
  $y^\prime$&-2.880&-1.783&-0.876&-0.491&-0.844&-1.019&-0.922&-1.318&-0.386 \\
$\lambda_1$&0.559&0.387&0.311&0.456&0.616&0.456&0.616&0.559&0.311 \\
$\lambda_2$&0.116&0.123&0.141&0.204&0.099&0.204&0.099&0.116&0.141 \\
$\lambda_3$&0.093&0.945&0.731&0.828&0.665&0.828&0.665&0.093&0.731 \\
$\lambda_4$&-0.101&-0.141&-0.702&0.168&-0.365&0.168&-0.365&-0.101&-0.702 \\
$\lambda_5$&-0.074&-0.884&-0.099&-0.968&-0.402&-0.968&-0.402&-0.074&-0.099 \\
$\lambda_S$&0.924&0.867&0.910&0.352&0.600&0.352&0.600&0.924&0.910 \\
$\eta_2$&0.182&0.158&0.156&0.226&0.173&0.226&0.173&0.182&0.156 \\
$\mu_1/\mathrm{GeV}$&41.982&42.672&46.821&102.293&22.346&102.293&22.346&41.982&46.821 \\
$\mu_2/\mathrm{GeV}$&71.135&69.551&72.458&94.039&66.776&94.039&66.776&71.135&72.458 \\
$\mu_S/\mathrm{GeV}$&98.904&94.206&91.722&95.987&99.757&95.987&99.757&98.904&91.722 \\
$v_1/\mathrm{GeV}$&102.562&119.174&129.687&142.845&96.180&142.845&96.180&102.562&129.687 \\
$v_2/\mathrm{GeV}$&223.600&215.206&209.039&200.278&226.419&200.278&226.419&223.600&209.039 \\
$v_S/\mathrm{GeV}$&26.778&42.620&41.949&19.861&42.840&19.861&42.840&26.778&41.949 \\
$m_{H^0}/\mathrm{GeV}$&90.749&88.633&89.541&139.698&86.112&139.698&86.112&90.749&89.541 \\
$m_{S^0}/\mathrm{GeV}$&28.652&45.074&47.940&8.761&26.970&8.761&26.970&28.652&47.940 \\
$c_{23}$&0.977&0.934&0.954&0.999&0.916&0.999&0.916&0.977&0.954 \\
$c_{13}$&0.994&0.985&0.985&0.995&0.985&0.995&0.985&0.994&0.985 \\
$c_{12}$&0.700&0.661&0.588&0.484&0.723&0.484&0.723&0.700&0.588 \\
  \bottomrule
\end{tabular}
\caption{ Exemplary parameter sets used in Fig. \ref{fig:constraintshiggs}.  }
\label{tab:amupars}
\end{table}
\end{landscape}


\end{document}